\documentclass[fleqn,usenatbib]{mnras}

\usepackage{newtxtext,newtxmath}

\usepackage[T1]{fontenc}
\usepackage{ae,aecompl}

\usepackage{graphicx}	
\usepackage{amsmath}	
\usepackage{amssymb}	
\usepackage{hyperref}

\title[IR burst in 4U 1728-34]{Discovery of a thermonuclear Type I X-ray  burst in infrared: new limits on the orbital period of 4U 1728-34}

\author[F. M. Vincentelli et al.]{
F. M. Vincentelli $^{1,3}$\thanks{F.M.Vincentelli@soton.ac.uk},
Y. Cavecchi $^{2}$,
P. Casella $^{3}$,
S. Migliari $^{4,5}$,
D. Altamirano $^{1}$,
\newauthor{~T. Belloni $^{6}$,
M. Diaz-Trigo $^{7}$
}
\\
$^{1}$Department of Physics and Astronomy, University of Southampton, SO17 1BJ, UK\\
$^{2}$Mathematical Sciences and STAG Research Centre, University of Southampton, SO17 1BJ, UK\\
$^{3}$INAF, Osservatorio Astronomico di Roma 
Via Frascati 33, I-00078 Monteporzio Catone, Italy\\
$^4$XMM-Newton Science Operation Center, ESAC/ESA, Camino Bajo del Castillo s/n, Urb. Villafranca del Castillo,\\
~28691 Villanueva de la Ca\~{n}ada, Madrid, Spain\\
$^5$Institute of Cosmos Sciences, University of Barcelona, Mart\'i i Franqu\`es 1, 08028 Barcelona, Spain\\
$^6$INAF, Osservatorio Astronomico di Brera Merate, via E. Bianchi 46, I-23807 Merate, Italy\\
$^7$ESO, Karl-Schwarzschild-Strasse 2, 85748 Garching bei M\"unchen, Germany
}

\date{Accepted XXX. Received YYY; in original form ZZZ}

\pubyear{2019}

\begin{document}
\label{firstpage}
\pagerange{\pageref{firstpage}--\pageref{lastpage}}
\maketitle

\begin{abstract}
We report the detection of an infrared burst lagging a thermonuclear Type I X-ray burst from the accreting neutron star 4U 1728-34 (GX 354-0). Observations were performed simultaneously with XMM-Newton (0.7-12 keV), NuSTAR (3-79 keV) and HAWK-I@VLT (2.2$\mu$m). We measure a lag  of  $4.75 \pm 0.5$ s between the peaks of the emission in the two bands. Due to the length of the lag and the shape of the IR burst, we found that the most plausible cause for such a large delay is reprocessing of the Type I burst X-rays by the companion star. The inferred distance between the neutron star and the companion can be used to constrain the orbital period of the system, which we find to be larger than $\sim$ 66 minutes (or even $\gtrsim$ 2 hours, for a realistic inclination $< 75^\circ$). This is much larger than the current tentatively estimated period of $\sim 11$ minutes.   We discuss the physical implications on the nature of the binary and conclude that most likely the companion of 4U 1728-34 is a helium star.
\end{abstract}





\section{Introduction} \label{sec:intro}
 
4U 1728-34 (GX 354-0) is one of the most studied neutron star (NS) low mass X-ray binaries (LMXBs) and shows the classical observational characteristics of this kind of weakly magnetized sources. The X-ray emission is known to show a quasi-periodic state transition (every $\sim$ 40 days) between a soft state and a hard state \citep{munios-darias}. In the first case the X-ray spectrum shows a thermal component modelled with a thin accretion disc \citep{shakurasunayev}; in the latter case the spectrum can be fitted with a power-law component with an high-energy cut-off, and it is usually explained in terms of inverse Comptonization by a corona of hot electrons close to the NS surface \citep{ng2010,egron2011,wang2019}.  On shorter timescales the source presents strong aperiodic variability, quasi-periodic oscillations \citep{disalvo2001}, and thermonuclear Type I X-ray bursts \citep[and \citealt{Strohmayer1996} for a review]{hoffman1976,basinska1984}. 
 
Thermonuclear Type I X-ray bursts (Type I X-ray bursts from now on) are sudden flashes observed from accreting NSs in LMXBs \citep{lewin1993}. They are usually observed in X-rays where the luminosity can reach up to the Eddington limit in few seconds. During the burst the X-ray spectrum is consistent with a black-body spectrum which slowly cools down after the peak \citep{swank1977,hoffman1977b,hoffman1977a,hoffman1978}. These sudden flashes are thought to be the consequence of the ignition and unstable burning of the matter accreted on the NS from the companion \citep[e.g.][see \citealt{gallowaykeek2017} for a recent review]{fujimoto1981}. The flame spreads from the initial ignition location until it engulfs all the star surface in few seconds \citep[e.g.][]{cavecchi2019}. The time scales of the bursts and their recurrence times depend on many factors, among which the most important ones are the accretion rate and the accreted matter composition. These determine which nuclear reactions take place. In particular, the duration of the tail of the bursts depends significantly on the composition of the accreted matter \citep[e.g.][]{schatz2001}.

Type I X-ray bursts from 4U 1728-34 have a short duration and a very high $\alpha$ value (the ratio of the  persistent fluence between bursts over the burst fluence\footnote{The total energy emitted per unit area.}). Such features are expected for helium-only burning bursts \citep{galloway2010}. Also, assuming  a solar composition for the accreted matter, the typical burst recurrence time of $\sim$3 hr cannot be reproduced  without leaving some hydrogen left to burn during the bursts  \citep[see][]{misanovic2010,galloway2010}.  This seems to point towards almost pure helium accretion from an evolved companion. Moreover \citet{galloway2010} reported a possible detection of a periodic signal at ~11 minutes in the persistent X-ray emission between the bursts, using Chandra observations taken in June 2006. Given the helium burning nature of the bursts, it was proposed that 4U 1728-34 was an ultra-compact binary with a He white dwarf feeding the neutron star. However, analysis of other X-ray observations could not find further evidence for a periodic signal at $11$ minutes.

 Type I X-ray bursts have been often detected also at UV and optical wavelengths \citep{grindlay1978,mcclintock1979,pedersen1982,hynes2006}. The observed light curves of the bursts in these bands differ usually from the X-ray ones, showing, for example, broader shapes \citep{lawrence1983}. The general interpretation is that the X-ray light is reprocessed by the accretion disc and the donor star, giving rise rise to a burst at lower energies, with a range of delays \citep{cominsky1987,obrien}. 
 
\begin{figure}
\centering
\includegraphics[width=0.9\columnwidth]{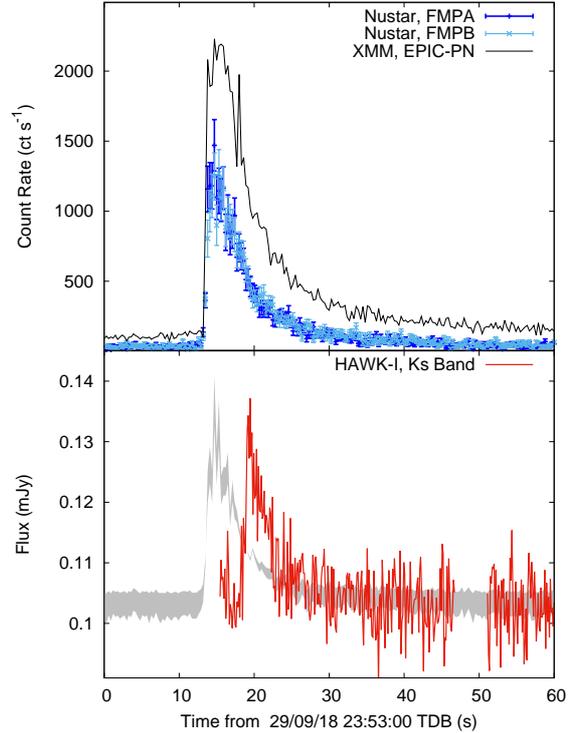}
\caption{Light curves of the strictly simultaneous multi-wavelength observations of 4U 1728-34 taken on 2018-09-29. Top panel: the X-ray burst is clearly detected by both \textit{NuSTAR} and \textit{XMM-Newton}. Bottom panel: the IR burst (red) superimposed on the re-scaled, \textit{NuSTAR} X-ray burst for comparison (grey).}
\label{fig:curve}
\end{figure}
\vspace{-0.3cm}

\section{Observations}

We performed a strictly simultaneous X-ray/IR observation of 4U 1727-34 {(R.A., DEC :  17:31:57.73 -33:50:02.5)} on 2018-09-29 (MJD  58420).  X-ray observations were done with \textit{XMM-Newton} and with \textit{NuSTAR}, while the IR band was covered by HAWK-I at Very Large Telescope (VLT).  {For all instruments, time was set to Dynamical Barycentric Time (TDB) system using the JPL Earth ephemeridis.} 

\vspace{-0.5cm}

\subsection{X-ray Observations}

\subsubsection{XMM-Newton}
We extracted data from the \textit{XMM-Newton} Epic-pn camera \citep{struder2001}.  The satellite observed continuously  between 20:39 and 02:44 UTC in Timing mode (Obs. ID 0824150601).  In order to extract the light curve we followed the procedure described in the SAS manual. In particular we selected the events  with PATTERN $<=4$, in the 0.7-10 keV energy range and using a box of $\approx86$ arcsecond of angular size (RAWX between 28 and 48). The event file was barycentred  using the command \textit{barycen}. The X-ray light curve was extracted with 60 ms time resolution and corrected from possible instrumental factors with the command \textit{epiclccorr}.

\subsubsection{NuSTAR}
Hard X-rays (3-79 keV) were collected with the two focusing telescopes on board of \textit{NuSTAR} \citep{harrison2013} from 22:13 to 08:49 UTC (Obs. ID 30401020010).  The light curve was extracted using the NuSTAR Data Analysis Software v1. We used the FTOOL \textit{nuproducts} from the focal plane modules A and B (FPMA and FPMB).  In particular we extracted events using a circle  of  100'' around the source between 3 and 10 keV. Events were binned in a light curve with 60 ms and barycentered  through the nuproducts pipeline.   
\vspace{-0.5cm}

\subsection{IR Data}

We collected IR  ($K_s$ band) high time resolution data with HAWK-I mounted at VLT UT-4/Yepun.   HAWK-I is a near-infrared wide-field imager   made up of four HAWAII 2RG 2048x2048 pixel detectors \citep{Pirard2004}.  Observations were taken from 23:53  to 00:53 UTC (Program ID 0101.D-0935). The observation was performed  in \textit{Fast-Phot} mode, reading only a stripe of 16 contiguous windows of 128x64 pixels in each quadrant. This allowed us to reach a time resolution of 0.125s. Every 250 exposures the data have a $\sim$3s long gap to transfer the data, emptying the instrument buffer. The instrument was pointed in order to put the source and a bright reference star (Ks $= 10.44$) in the top-right quadrant (Q3). Photometric data were extracted using the ULTRACAM data reduction software tools \citep{dhilon2007}. Parameters for the extraction were derived from the bright reference star, to which the position of the target was linked in each exposure. {To account for seeing effects, such as spurious  long term trends due to changes in the background, we took the ratio between the source and the reference star count rate was used}. The time of each frame was then put in the TDB system using the barycentering software developed by \citet{Eastman}.

\vspace{-0.3cm}

\section{Analysis and Results} \label{sec:style}

\subsection{Detection of an IR Burst}

 Both X-ray satellites detected a Type I burst at  23:53:13 TDB. Fig. \ref{fig:curve}  shows the strictly simultaneous light curves in the two bands (0.7-12 and 3-79 kev). The observations show the typical behaviour for this kind of phenomenon, with a steep rise followed by a slower decay.  HAWK-I observations started at  23:53:15.5 TDB. A sudden and sharp rise in the light curve, followed again by a slower decay is seen at 23:53:19. We measured a peak magnitude of Ks $= 14.23 \pm 0.03$ (not de-reddened flux of $\simeq$ 0.135 mJy). For the rest of the observation the light curve remained approximately constant at Ks $= 14.516 \pm 0.001$ ($\simeq$ 0.104 mJy). We computed the cross-correlation function (CCF) between the X-ray and IR light curves using  data from both satellites. Light curves were rebinned with a time bin of 0.5 s. The CCF was computed following the formalism described in \citet{gandhi2010}. While the XMM CCF has a maximum at 4.5$\pm$0.25 s, the NuSTAR one peaks at 5$\pm$0.25 s. {We investigated the origin of this marginal difference by using the same energy band for both satellites (3-10 keV). Given that the results did not change significantly in this case, we ascribe the origin of the marginally different peaks to the non-identical effective areas of the satellites coupled with the evolution of the X-ray spectrum during the burst. Pile-up effects could also marginally modify the burst profile observed by XMM. However, the two measurement of the lag are consistent, suggesting that also this effect is negligible.} We also note that the slightly asymmetric shaped CCF suggests that the characteristic decay-timescale of the IR burst is shorter than the X-ray one. An exponential fit ($A~e^{-t/\tau}+C$) gives $\tau_{\rm{IR}}=3.3\pm 0.4$s versus $\tau_{X}=4.4\pm 0.2$s: thus the difference in length is marginal.  We note however that the observed e-folding time $\tau$ can be affected by the level of the persistent emission, especially in the IR  case, as the amplitude of the peak is only $\approx30\%$ of the level of the persistent.

\begin{figure}
\centering
\includegraphics[width=0.9\columnwidth]{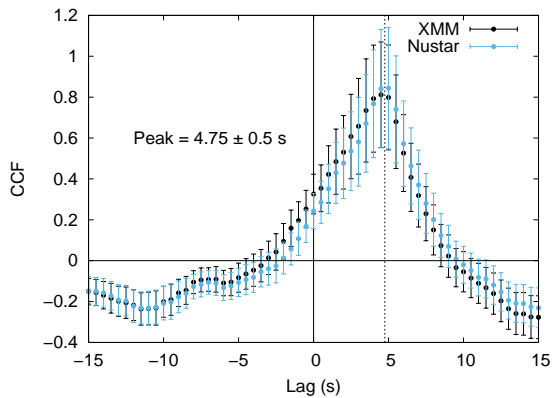}
\caption{IR (HAWK-I) vs X-ray (XMM in black, NuSTAR in blue)  cross-correlation function. The average peak is at 4.75 s.}
\label{fig:ccf}
\end{figure}

\vspace{-0.5cm}

\subsection{Orbital Period}
\label{section:period}

The detection of an IR burst and its lag with the X-ray one can be used to put geometrical constraints to the system. In particular, the Optical-IR counterparts of an X-ray burst are expected to arise from the reprocessed emission by the disc and the donor star \citep{obrien}.  The lag between the two lightcurves would then be due to the difference in path between the direct burst emission and the reprocessed light.  In the following, we will assume an average lag of $4.75 \pm0.5$s, noting that the results presented here are insensitive to small changes of this value. Converting the measured lag in light travel-time distance, we obtain a distance of $\simeq10^6$ km. Such large distance could still arise from both the outer edge of the disc or from the companion star.  We note, however, that past observations of optical bursts show always a rise starting almost simultaneously to the X-ray burst \citep{hynes2006burst,paul2012}, as expected from the response of a disc \citep{obrien}. On the contrary, we observe the IR rise taking place a few seconds after the X-rays. {We also notice that, contrarily to what is expected from a burst reprocessed by an extended disc, the length of the reprocessed burst is comparable with the X-ray one.  We conclude that the IR burst arises most probably from the companion star, and not from the disc.}

Under this hypothesis, the measured lag can be interpreted as the light-travel time difference between the two bodies of the system, and it can be therefore used to  constrain the period of the binary. The conversion into distance, however, depends  on the orbital phase and the inclination. 
Assuming a circular orbit, and neglecting the size of the companion star, the dependence of the lag on the orbital phase is the following  \citep{obrien}:
\begin{equation}
\centering
    \tau = \frac{a}{c} (1 + \sin i \cos \phi)
    \label{equ:lag}
\end{equation}
Where $a$ is the binary separation, $i$ is the inclination, and $\phi$ is the orbital phase\footnote{We adopt the convention that at phase $\phi=\pi$ the companion is between the neutron star and the observer.}. This means that if the system is face-on ($i=0^\circ$), the measured lag would coincide (in light travel time) with the semi-major axis. 
For inclinations $> 0^\circ$, the measured lag would correspond to a range of possible binary separations, depending on the orbital phase at the time of the burst.

In the most extreme case of $i=90^\circ$ and $\phi=0$ (i.e., if the system is ``edge-on'' and the companion is behind the NS), the actual orbital separation (in light travel time) would be  half the measured X-ray/IR lag: this is the smallest $a$ allowed for the system. Given that the period $P \propto a^{3/2}$, this sets a lower limit to the orbital period. In order to be conservative, we also take into account the possible extra delay due to the reprocessing time. It has been shown that the typical reprocessing time of an X-ray type I burst illuminating a star is  $\leq 0.2 $ s \citep{cominsky1987}. This means that the minimum orbital separation for a 4.75 s lag will be given (in light travel time) by half the measured lag minus the maximum reprocessing time. To be conservative we set such distance as 2 light seconds ($a_{min}/ c= 2 s$).

\begin{figure}
\centering
\includegraphics[width=0.9\columnwidth]{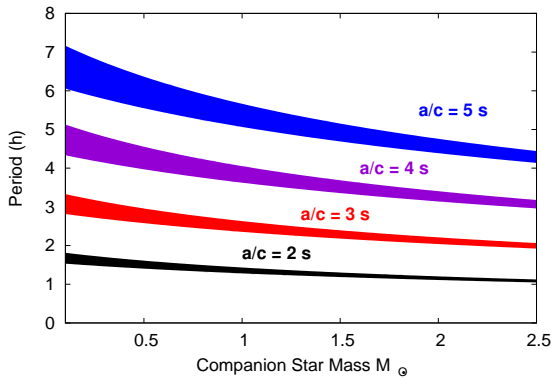}
\caption{Expected periods for a binary system with a neutron star of a mass between 1.4 and 2 M$\odot$   as a function of the mass of the donor star. The five curves correspond to five different orbital separations (from bottom to top:   2, 3, 4 and 5 light seconds).}
\label{fig:period}
\end{figure}

Through the Keplerian law we then computed the orbital period. Fig. \ref{fig:period} shows the orbital periods as a function of the companion mass for different orbital separations (in light-travel time) and for neutron star masses ranging from 1.4 and 2 M$_\odot$ \footnote{$P\propto M_{tot}^{-1/2}$: higher $M_{NS}$ will give lower orbital periods.}. We find that also in the extreme case of a 2 $M_\odot$ NS (lower limit of the filled curves), the orbital period of the system will always be $\gtrsim$ 1.1 h ($\approx$66 min). We note that this limit is extremely conservative. For example, X-ray spectral measurements indicate that the inclination should be between 23$^\circ$ and 53$^\circ$ \citep{wang2019}, while the fact that 4U1728-34 does not show dips or eclipses, implies that the inclination must be $\lesssim 75^\circ$ \citep{frank1987}. For instance at $i=75^\circ$ the orbital separation would be $\approx  2.5$ light seconds, which means that we do not expect a period $\lesssim$ 2 hr. {We also note that this limit is still valid even in the case the reprocessed burst originated from the outer edge of the disc: the orbital separation (and therefore the orbital period) in this scenario would be in any case larger than the one measured with our assumptions. }

\vspace{-0.5cm}

\subsection{Preliminary Modelling}

\label{section:mod}
\begin{figure*}
\centering
\includegraphics[width=0.9\textwidth]{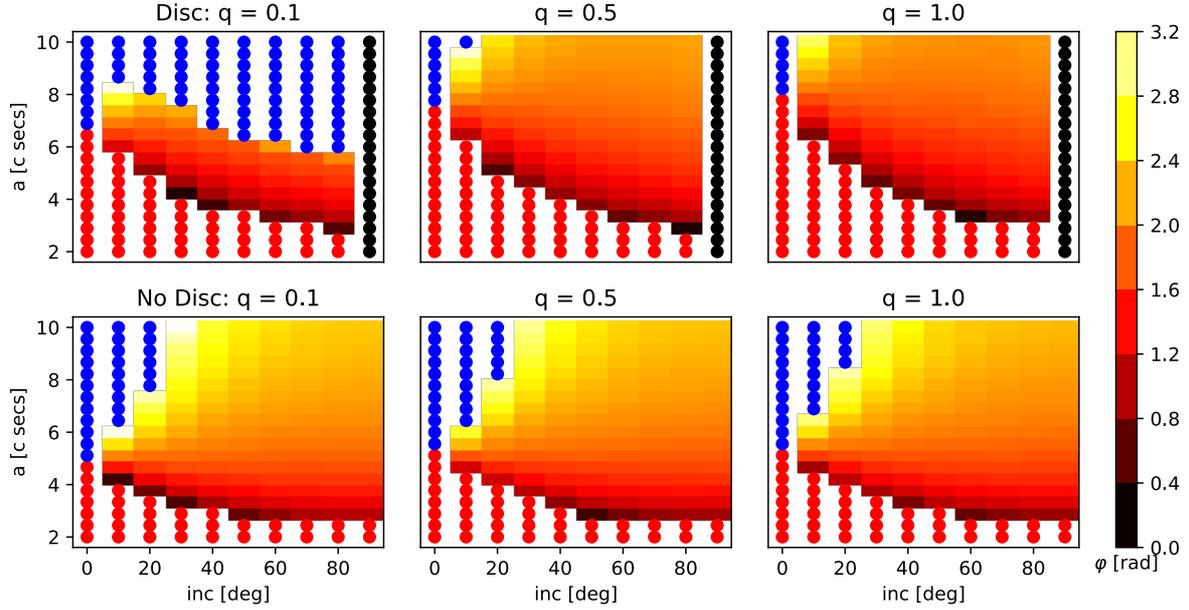}
\caption{Phases at which the observed lag between X-ray and IR is $4.75$ s for the case including a disc and the case without. The color of the surfaces corresponds to the orbital phase, as indicated by the color bar. Red dots mean no possible phase, all lags $< 4.75$ s, blue dots mean no possible phase, all lags $> 4.75$ s. Black dots mean no Type I burst detected in X-ray (because the disc is eclipsing the NS).  Some parameter combinations lead to an IR profile with two comparable peaks due to the disc response. Using the first of these peaks would not be consistent with the observations and therefore we exclude these cases.
}
\label{fig:lag}
\end{figure*}
 
The considerations discussed in Sec. \ref{section:period} and Eq. (\ref{equ:lag}) neglect the finite size of the companion and the presence of the disc. To include these we have written a code to calculate synthetic light curves of bursts in X-rays and IR. We followed \citet{obrien} and the references therein. Our code includes the NS, assumed to be spherical, a disc and the companion, assumed to be in circular orbit and filling its Roche lobe. We neglect the stream from the lagrangian point L1 to the disc. For the disc height we use $H \propto (R / R_{\rm{out}})^\beta$, so that the disc can have a finite thickness at the star surface. {The data available do not allow a full fit of the parameters. So we run exploratory models, under some reasonable assumptions, in order to constrain the lowest orbital period. The outer edge thickness is $H_{\rm{out}} = 0.016 * R_{\rm{out}}$. We set $\beta = 1.1$ and the outer radius $R_{\rm{out}}$ is given by the circularization radius  \citep[see e.g.][]{maccarone2013}. The code follows \citet{obrien} in accounting for the illumination and heating of the companion and disc, and computes directly the observed light curves summing the emission from all the components. The base temperature of the companion is set to $T = 5.5 \times 10^3$ K, based on the persistent IR emission \citep{marti}. }

We ran two sets of simulations: including the disc or not. We explored a range of $19$ values for the orbital separation $a$ from $2$ to $10$ light seconds, a range in mass ratio $q$ of $10$ values from $0.1$ to $1$ and $10$ values of inclination $i$ from $0$ to $90$ degrees. For each combination of $a$, $q$ and $i$ we sampled $10$ orbital phases between $0$ and $\pi$. {In  Fig. \ref{fig:lag} we show the orbital phase that would give a lag of $4.75$ s as a function of $a$, $i$ and $q$}. If we do not include the disc, once we account for the finite size of the Roche lobe using $a' = a - R_{\rm{L}}$, the delays are well described by Eq. (\ref{equ:lag}).   $R_{\rm{L}}$ is the equivalent Roche lobe radius: the radius giving the volume of the Roche Lobe centered on the companion. An approximate formula is \citep{eggleton1983}:
\begin{equation}
\centering
    R_{\rm{L}} = a \frac{ 0.49 q^\frac{2}{3}}{0.6 q^\frac{2}{3} + \ln(1 + q^\frac{1}{3})}
\label{eq:reql}
\end{equation}
where the mass ratio $q = M_{\rm{c}} / M_{\rm{NS}}$. When the disc is included, the phase space where a solution is possible is heavily reduced. One important reason for this is that the disc obscures part of the companion. We note that we did not find solutions for $a \lesssim 2.9$ light seconds within the parameter space we explored here. For $q = 1$, the minimum $a$ is $\sim 3.3$ light seconds when the disc is accounted for, suggesting that a white dwarf (WD) on 11 minutes orbit is unlikely. 
 
 \vspace{-0.3cm}

\section{Discussion and Conclusions}

We discovered an IR burst taking place  4.75 s after an X-ray Type I burst in the accreting NS 4U 1728-34.  Due to the length of the lag, and the similar duration of the two bursts,  we interpret the IR burst as reprocessing from the companion star. Under this hypothesis, we found that the orbital period of 4U 1728-34 must be greater than $\sim$ 1.1 hours. Such an estimate is in contradiction with the proposed ultra-compact nature of the system and the implication that the companion must be an evolved star, in particular a WD.
 
The main arguments in favour of an evolved donor star in a short binary system were: the evidence of hydrogen poor Type I bursts; the marginal detection of a periodicity of $\sim 11$ minutes \citep{galloway2010}; and the fact that the mass transfer rate estimated for 4U 1728-34 suggests a period of $\sim$20 minutes according to the period-$\dot{M}$ correlation observed in the systems with an evolved donor \citep{heinke2013}.

{Comparing the radius of a WD from its mass-radius relation to a Roche lobe radius with $2s \lesssim a / c \lesssim 4s $,  we found that to have a WD as a donor star, it has to have a mass of a few $10^{-3} \rm{M}_\odot$. Our modelling indicates that the parameter space allowed for low $q$ becomes very small and for $q\sim10^{-3}$ only a neglible part of the parameter space is allowed. Such low $q$ NS-WD system would also have an accretion rate of order $10^{-13} \rm{M}_\odot / \rm{yr}$ or less \citep[see e.g. ][]{deloye} , which is much lower than the reported accretion rates  \citep{galloway2010,heinke2013}.  All these considerations imply that a WD companion seems unlikely.}

On the other hand, hydrogen-poor bursts could also occur if the donor is a helium star. These systems are believed to originate from Be-Xray binaries and thought to evolve into WD-NS systems \citep{dewi2002}.  Numerical simulations have shown that during their evolution, these systems can reach orbital periods of few hours \citep{dewi2002} and therefore would be consistent with our lag measurement. 

In conclusion, we note that  this result is based only on one single detection, thus we can place only a lower limit to the orbital period. More observations of this kind, with several measurements as a function of the orbital phase can be used to obtain a precise estimate of the orbital parameters.  

\emph{Acknowledgements}: This Letter benefited from the meeting `Looking at the disc-jet coupling from different angles' held at the International Space Science Institute in Bern, Switzerland.  FV thanks Omer Blaes and Soton ``binary group''  for useful discussions.  FV acknowledges support from STFC under grant ST/R000638/1. YC is supported by the EC Marie Sk\l{}odowska-Curie Global Fellowship grant No. 703916.  FMV acknowledges support from STFC under grant ST/R000638/1. DA is supported by the Royal Society. TMB and PC acknowledges financial contribution from the agreement ASI-INAF n.2017-14-H.0


\vspace{-0.7cm}
\bibliographystyle{mnras}
\bibliography{bib} 

\bsp	

\label{lastpage}
\end{document}